\documentclass[aip,rsi,amsmath,amssymb,reprint]{revtex4-1}
\usepackage{amssymb}
\usepackage{amsmath}
\usepackage{graphicx}
\usepackage{dcolumn}
\usepackage{bm}
\usepackage{docs}
\usepackage{url}
\usepackage[pdftex,colorlinks=true,	linkcolor=red,	citecolor=blue,	urlcolor=blue]{hyperref}

\begin{document}


\title{A novel time-of-flight mass spectrometer using radial extraction from a linear quadrupole trap for atomic, molecular, and chemical physics}

\author{Steven J. Schowalter}
\email{schowalt@physics.ucla.edu.}
\author{Kuang Chen}
\author{Wade G. Rellergert}
\author{Scott T. Sullivan}
\author{Eric R. Hudson}
\affiliation{Department of Physics and Astronomy, University of California, Los Angeles, California 90095, USA}
\date{\today}

\begin{abstract}
We demonstrate the implementation of a simple time-of-flight (ToF) mass spectrometer with medium-mass resolution ($m/\Delta m\sim50$) geared towards the demands of atomic, molecular, and chemical physics experiments.  By utilizing a novel radial ion extraction scheme from a linear quadrupole trap, a device with large trap capacity and high optical access is realized without sacrificing mass resolution.  Here we describe the construction and implementation of the device as well as present representative ToF spectra.  We conclude by demonstrating the flexibility of the device with proof-of-principle experiments that include the observation of molecular-ion photodissociation and the measurement of trapped-ion chemical reaction rates. 

\end{abstract}

\maketitle

\section{Introduction}

In atomic, molecular, and chemical physics, linear quadrupole traps (LQTs) are ubiquitous. They are routinely used in the study of topics ranging from quantum information\cite{Campbell2010,Hume2007,Haffner2005} to plasma dynamics\cite{Drewsen1998} to precision measurement\cite{Chou2010,Rosenband2008,Koelemeij2007}. Moreover, in the rapidly emerging field of hybrid atom-ion devices, LQTs are now being employed to study the interactions of cold atomic and molecular ions with neutral atoms and molecules\cite{Zipkes2010,Schmid2010}. These studies have already resulted in important quantum chemistry measurements\cite{Rellergert2011,Hall2011,Grier2009} and promise to enable new technologies and science such as the implementation of novel quantum gates\cite{Idziaszek2007}, probes of quantum gases\cite{Kollath2007}, observation of novel charge-transport dynamics\cite{Cote2000}, and the sympathetic cooling of atomic and molecular systems which cannot be laser cooled\cite{Hudson2009,Smith2005}. 

While the LQT is a thoroughly proven device, direct identification of non-fluorescing ions in a LQT is not straightforward and is often problematic.  Current methods to identify ions in a LQT include nondestructive resonant-excitation mass spectrometry using laser-induced fluorescence\cite{Baba1996,Roth2006}, destructive resonant-excitation mass spectrometry using channel electron multiplier (CEM) detection\cite{Sullivan2011,Offenberg2009,Douglas2005}, LQT mass-filtering techniques\cite{Ostendorf2006,Douglas2005}, and molecular dynamics (MD) simulations\cite{Schiller2003,Roth2008}. The first three methods are complicated by LQT properties which give rise to nonlinear resonances, multiple secular frequencies, and trap-depth limitations. These complications may lead to either the incorrect labeling or the lack of observation of certain ions.  The last method is limited by the inability to even indirectly identify the non-laser-cooled ion.  These limitations and complications can be avoided if ions are instead directly observed in a simple, straightforward manner using time-of-flight (ToF) mass spectrometry.

\begin{figure}[b]
\resizebox{\columnwidth}{!}{
    \includegraphics{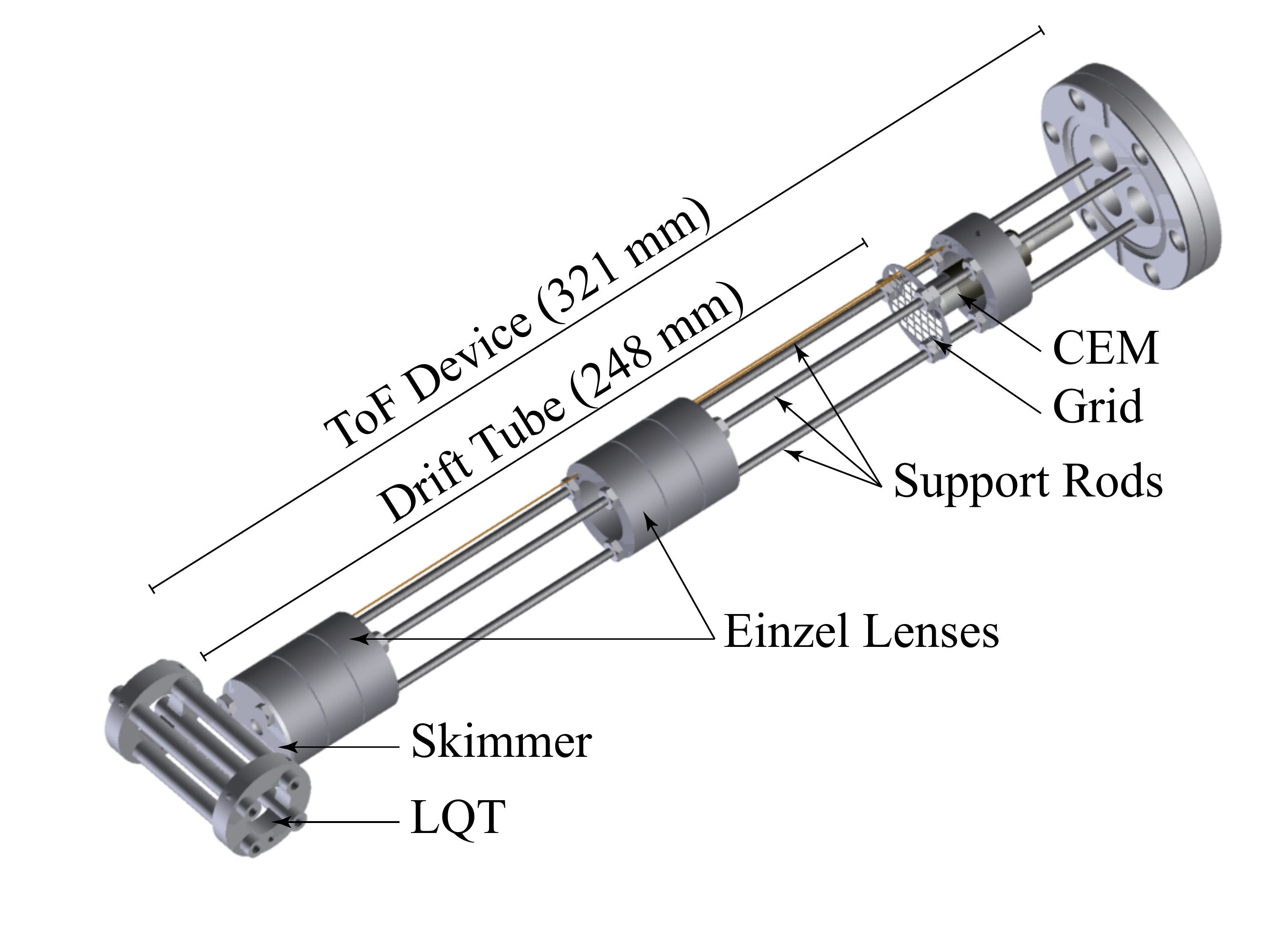}
}  \caption{(Color online) A 3D model of the LQT-ToF device.   \label{fig:app}}
\end{figure}

Unfortunately, while commercial devices which couple ion traps with ToF units do exist, they do not offer a practical solution for most experimental architectures employed in atomic, molecular, and chemical physics.  First, commercial ToF devices are predominantly geared towards life science research where the study of large molecules requires extremely high mass resolutions ($m/\Delta m = 1000 -100000$)\cite{Makarov2000,Tanaka1988}. To meet these stringent demands, commercial ToF devices are large, highly integrated designs that are constructed to be the principle feature of the experimental apparatus. Second, in most commercial units the ToF unit is coupled to a 3D Paul trap rather than a LQT\cite{Douglas2005}.  While this type of device works well for many applications, 3D Paul traps, unlike LQTs, are not amenable to studies that require large trap capacities and high optical access to the trapped ions -- demands typical of atomic, molecular, and chemical physics experiments.  And finally, though some commercial devices incorporate a LQT, the trapped ions are typically axially ejected from the LQT and subsequently accelerated orthogonally into the ToF drift tube to recover good mass resolution\cite{Dawson1989,Campbell1998}. Although this is an effective coupling technique, the commercial implementation of orthogonal acceleration is technically complicated, space-demanding, and expensive. 

Here we describe a simple LQT-ToF device, which offers near-isotopic mass resolution for atomic and molecular ions with masses of a few 100~amu. The straightforward, compact design ensures that the device does not impede non-ToF related functions by limiting optical or spatial access, making it ideal for most atomic, molecular, and chemical physics applications. This is accomplished by coupling the LQT to the ToF apparatus through a novel radial extraction method resulting in good mass resolution and without the complications associated with axial ejection and orthogonal acceleration. Additionally, due to the use of ion optics to enhance ion detection efficiency, long interrogation times do not need to be sacrificed for higher measurement rates, as in most commercial devices, in order to build statistics on reasonable time scales.  Finally, the proposed device can be built for a small fraction of the cost of current commercial ToF devices.

In the remainder of this article, we demonstrate the implementation and characterization of the LQT-ToF device.  We first discuss the design and construction of the device addressing the ToF unit, the LQT used to trap and interrogate a sample of ions, and the associated electronics. Secondly, we describe the calibration of the device using mass-filtering techniques. We conclude with proof-of-principle experiments which demonstrate the effectiveness of the device as a simple, practical method of mass spectrometry for use in atomic, molecular, and chemical physics.

\section{Experimental Design}
The ToF device, shown in Fig. \ref{fig:app}, is built onto a $2$-$3/4$'' ConFlat flange and is supported by three $\#4$-$40$ threaded rods. The $248$~mm drift tube, confined by a surrounding $2$-$3/4$'' ConFlat nipple, includes ion optics which are used to guide ions into the detector region to increase the detection efficiency of the device.  The detector region is composed of a CEM held at $-1600$~V and a stainless steel grid used to ensure that the drift tube remains field-free, apart from the field created by the ion optics. The total length of the device is $321$~mm and is designed to couple with a LQT ($r_0=7.92$~mm and $r_e/r_0=0.401$ where $r_0$ and $r_e$ are the field and electrode radii, respectively) via pulsed radial extraction. The entire apparatus is maintained at background pressures below $10^{-8}$~mbar by a triode sputter-ion pump.

\subsection{Radial Extraction}\label{radial extraction}

\begin{figure}[b]
\resizebox{\columnwidth}{!}{
    \includegraphics{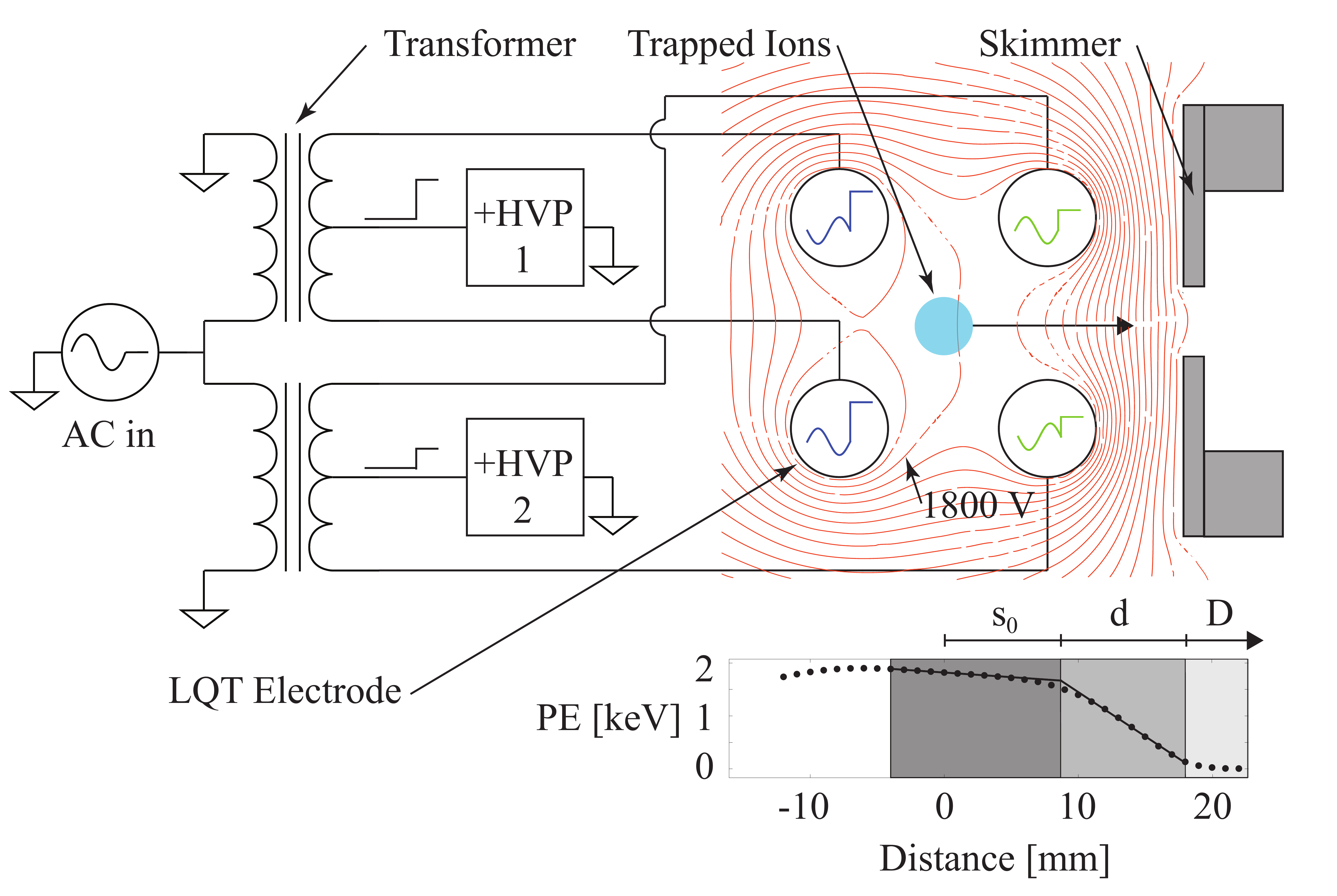}
}  \caption{(Color online) A rf  signal is applied to two center-tapped transformers to create the quadrupole trapping potential.  After the trapping period the primary rf is switched OFF and the +HV pulsers (+HVPs) are switched ON, ejecting the ions from the LQT. The two-stage acceleration scheme is shown by the equipotential contours ($100$~V spacing) and the on-axis potential energy plot.\label{fig:trap}}
\end{figure}
 
To simplify the coupling process, the ToF device couples directly to a LQT through radial extraction.  This method is superior to that of the normal axial extraction for three critical reasons.  First, field lines from the axially-confining end caps do not penetrate far into the trapping region. If the ions are pulsed from the trap axially (\emph{i.e.} by grounding an end cap), relatively long durations are required to completely empty the trap due to this weak coupling. This leads to a relatively large temporal spread of ions, which results in poor mass resolution.  This problem can be partially solved by using segmented trap electrodes\cite{Wilcox2002} to apply a more strongly-coupled, axially-ejecting field to the trapped ions; however, this solution requires more sophisticated machining and electronics. Second, the axial distribution of trapped ions in a LQT is generally larger than the radial distribution.  This leads to a larger distribution of initial energies and overall flight distances, and thereby poorer mass resolution when ions are axially pulsed. Third, it is often the case in atomic, molecular, and chemical physics that the LQT axis is used for the introduction of cooling or spectroscopy beams, which generally precludes axial extraction into a ToF device. 

Despite the obvious drawbacks of axial ion extraction, there have been relatively few implementations of radial ion extraction into a ToF device. In the patent of Franzen et al.\cite{Franzen1998}, the use of plate electrodes external to the LQT is proposed to apply pulsed potentials that radially eject the ions into the drift tube.  These plates, however, obstruct optical access to the trap and also require space in the vacuum chamber.  Additionally, field lines from the external plates do not penetrate far into the LQT, resulting in weak coupling between the plates and the trapped ions.  As a result it is difficult to establish an optimal ion extraction scheme using this method and no experimental data is found in the literature.  Unlike the design proposed by Franzen, radial extraction in our device is achieved by applying the extraction pulses directly to the trap electrodes, in a manner similar to that proposed by Makarov in Ref. \onlinecite{Makarov2009}, eliminating the need for external plate electrodes.  

For the operation of our device, various ions are introduced to the trap by ablating solid precursors with a pulsed Nd:YAG laser as in Ref. \onlinecite{Chen2011}.  For optimal loading, the radiofrequency (rf) trapping voltage is applied 50~$\mu$s after the ablation pulse.  This rf is applied to all four electrodes with the appropriate phases to create a quadrupole trapping potential with typical values of $\Omega\sim2\pi\times500$~kHz and $V_{\mathrm{rf}}\sim100$~V leading to Mathieu $q$ parameters of roughly $0.4$ for $m/Q\sim150$~amu/e, where  $\Omega$ is the trapping frequency,  $V_{\mathrm{rf}}$ is the rf amplitude, $q=\frac{4QV_{\mathrm{rf}}}{mr_0^2\Omega^2}$, and $m/Q$ is the mass-to-charge ratio of the trapped ion. To be clear, the prefactor of $4$ in the expression for $q$ is a result of applying voltage to all four electrodes rather than to only a diagonal pair.

After the trapping period, the rf is switched off and the ions are briefly allowed to expand for $\sim1.5$~$\mu$s. This delay redistributes the ions such that higher energy ions will either receive a lower extraction energy or have to travel farther before reaching the detector, leading to improved mass resolution\cite{Wiley1955}. To realize maximum mass resolution a two-stage acceleration scheme is implemented following the delay.  Extraction potentials of roughly $2000$~V (blue) and $1800$~V (green), as shown in Fig. \ref{fig:trap}, are applied to vertically paired electrodes.  This results in the radial extraction of ions from the trapping region, through the right pair of electrodes, and into the ToF drift tube which is referenced to ground. The electronics used to provide the pulsing voltages include two sets of positive-high-voltage power supplies and pulsers.  A circuit, shown schematically in Fig. \ref{fig:trap}, is used to combine the pulsing and trapping potentials.  Amplified rf voltage is applied to the primary windings of two hand-wound center-tapped transformers.  The center-taps of the secondary coils are connected to the high-voltage pulsing units and the two pairs of secondary wires are connected to vertically paired trap electrodes.  When the trapping rf is on, the pulser outputs are grounded, resulting in the normal quadrupole trapping potential.  However, when the trapping rf is off and the pulser outputs are high, identical high-voltage is applied to the vertically paired electrodes, resulting in the radial ejection of the ions into the ToF drift tube. A rf switch with a $\sim5$~$\mu$s fall-off time is used to switch off the rf prior to pulsing.  This large fall-off time relative to the pulse delay ($\sim1.5$~$\mu$s) guarantees the remainder of an rf component in the pulsed high voltage.  The phase of this rf can be tuned to optimize ion detection, mass resolution, and signal stability which may be otherwise limited due to fringing fields, charging effects, or engineering imperfections in the experiment.

To optimally implement the two-stage acceleration of the ions from the trap, LQT-ToF dimensions are determined based on the work of Wiley et al in Ref. \onlinecite{Wiley1955}.  Critical distances, shown in Fig. \ref{fig:trap}, include the length of the drift tube, $D$, the mean length of the first acceleration stage, $s_0$, and the length of the second acceleration stage, $d$. The extraction voltages ($2000$ and $1800$~V) establish the equipotential contours shown in Fig. \ref{fig:trap}. The corresponding potential energy plot shows the approximately uniform two-stage extraction field along the extraction axis.  The first stage (dark grey) has a mean length of $s_0=8.71$~mm and field of $E_s=174$~V/cm.  The second stage (medium grey) has a length of $d=9.29$~mm and field of $E_d=1671$~V/cm.  Due to the grounded skimmer and grid, a field less than $20$~V/cm extends past $5$~mm into the drift tube (light grey). In this two-stage scheme, the distribution of total energies received by the ions is much smaller than in the single-stage scheme, which results in higher mass resolutions as originally asserted in Ref. \onlinecite{Wiley1955}.

\subsection{Ion Optics}
\begin{figure}[b]
\resizebox{\columnwidth}{!}{
    \includegraphics{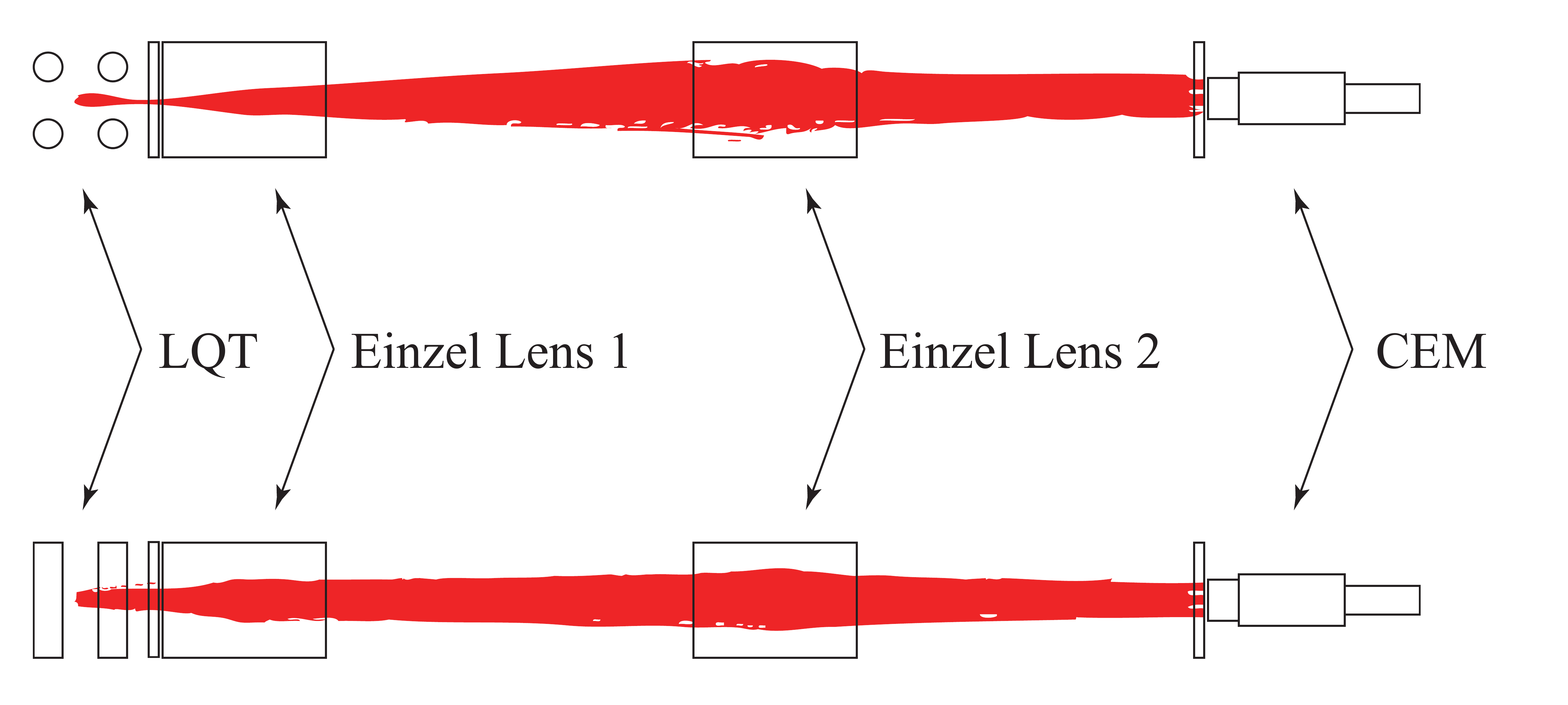}
}  \caption{(Color online) Simulated ion trajectories in the radial (top) and axial (bottom) plane of the LQT.  Ions in the radial plane have a higher divergence due to the focusing effect of the right lateral pair of LQT electrodes, motivating the use of two Einzel lenses to collimate the ion beam.  For clarity, only a subset of the axial ion distribution is shown. \label{fig:trajectories}}
\end{figure} 

In order to establish a large detection efficiency ion trajectories in both the radial and axial planes must be corrected to prevent ions from colliding with the walls.  This is complicated by the difference in divergences of ion trajectories in the radial and axial planes, which occurs as a result of the extraction geometries.  To correct this astigmatic ion beam we employ two Einzel lenses ($20$~mm diameter, $127$~mm center-lens length, $1$~mm lens gap) with different focal lengths. The first Einzel lens, added immediately after the skimmer, approximately collimates the ions in the axial plane and sufficiently decreases the divergence in the radial plane to avoid ion loss. A second Einzel lens, added closer to the detection region, slightly focuses ions in the axial plane while approximately collimating the ions in the radial plane. Setting the first and second Einzel lenses to $1300$ and $900$~V, respectively, leads to optimal detection by the CEM, as shown in Fig. \ref{fig:trajectories}.

While, the use of ion optics increases detection efficiency by enabling the detection of ions that would otherwise collide with the walls, those ions with bent trajectories travel farther, thus compromising mass resolution.  To mitigate this effect, a skimmer with a $5.6$~mm diameter aperture is used to block ions with extremely off-axis trajectories from entering the drift tube.

\section{Calibration of the Device}
\begin{figure}[t]
\resizebox{\columnwidth}{!}{
    \includegraphics{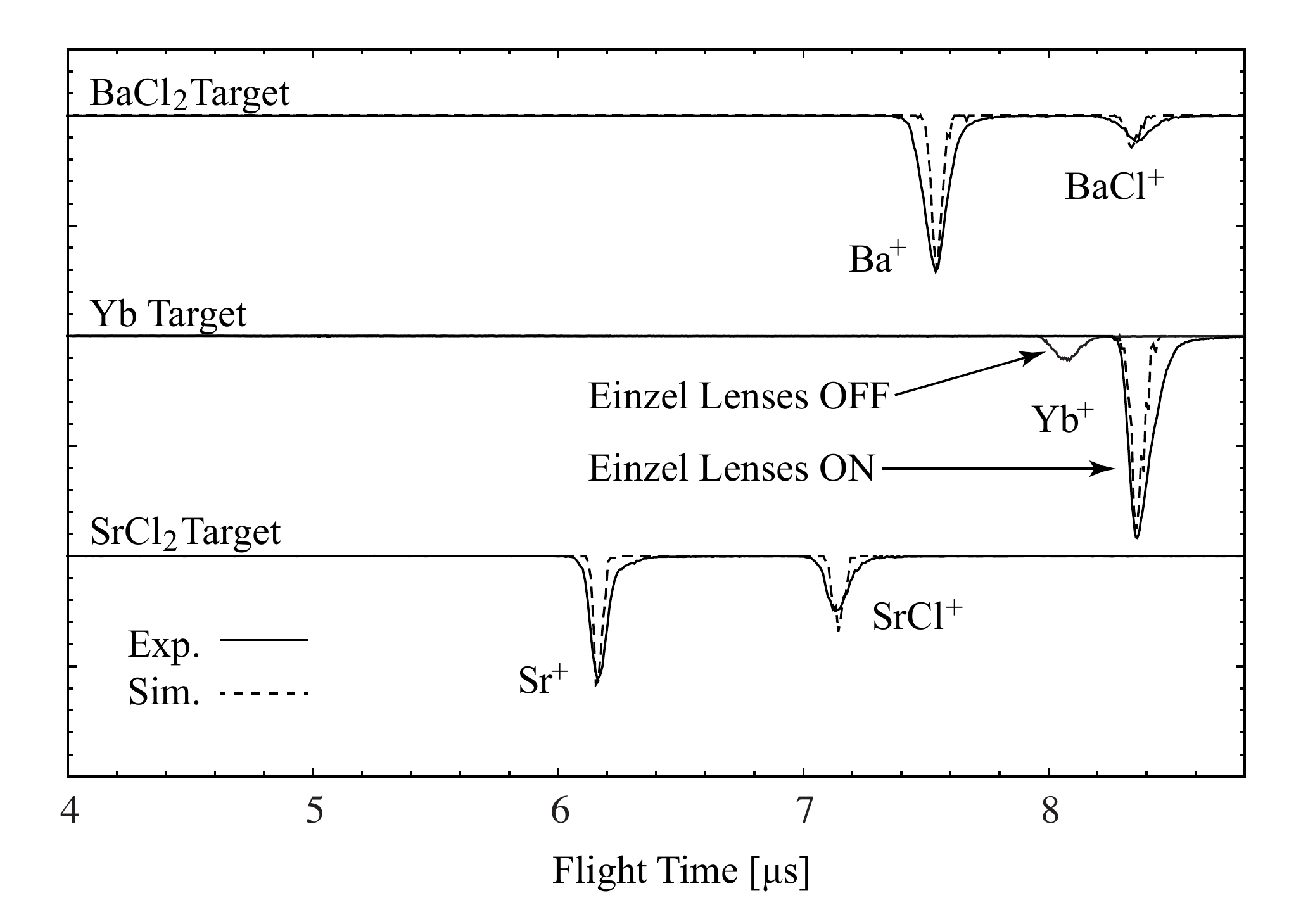}
}  \caption{Experimental (solid) and simulated (dashed) ToF spectra of trapped ions which are produced by the ablation of BaCl$_2$, Yb, and SrCl$_2$ solid targets.  The experimental mass resolution is shown to be slightly less than that of the simulation.  As shown in the Yb spectrum, the use of Einzel lenses increases ion detection by nearly an order of magnitude.  All other spectra include the use of Einzel lenses.  \label{fig:spectra}}
\end{figure} 

The device was calibrated and tested by trapping ions in the LQT and subsequently recording their ToF spectra.  Representative ToF spectra produced by ablating BaCl$_2$, SrCl$_2$, and Yb targets are shown in Fig. \ref{fig:spectra}. These targets were chosen due to the cold atomic and molecular ion community's interest in their ablation products, namely Ba$^+$\cite{Roth2005}, BaCl$^+$\cite{Hudson2009}, Sr$^+$\cite{Margolis2004,Leibrandt2009}, SrCl$^+$\cite{Hudson2009}, and Yb$^+$\cite{Rellergert2011,Zipkes2010}.  

\begin{figure}[b]
\resizebox{\columnwidth}{!}{
    \includegraphics{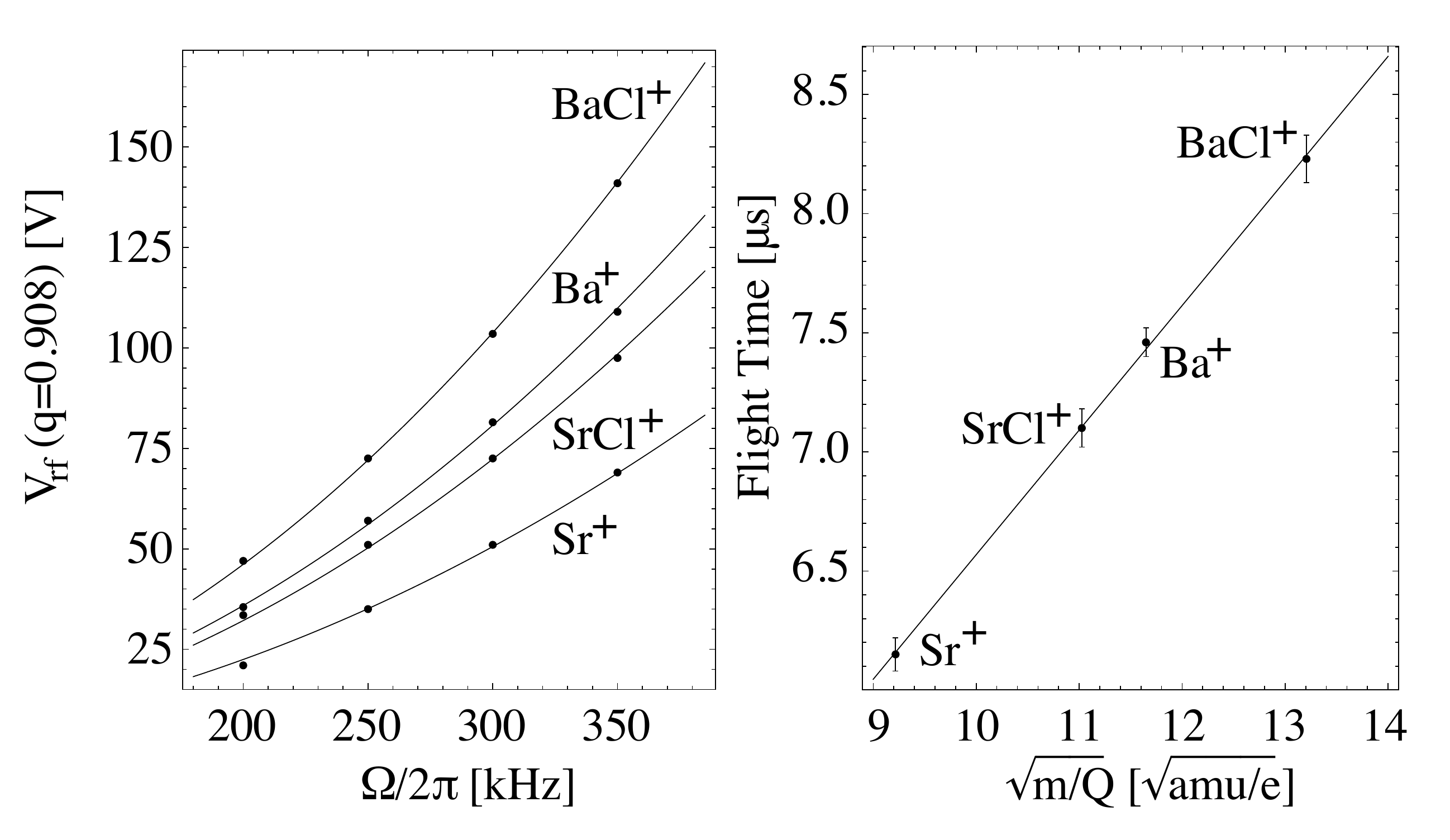}
}  \caption{The left panel shows the quadratic relationship between trap voltage and trap frequency for $q=0.908$, which is used to calculate $m/Q$ ratios for each observed ToF peak.  The right panel serves as the $m/Q$ ratio to flight-time calibration.   \label{fig:calibration}}
\end{figure}

To calibrate the ToF signal, the mass-to-charge ratio, $m/Q$, of each peak in the recorded ToF spectra was determined using standard mass-filtering techniques\cite{Douglas2005}. For each $m/Q$ peak, the LQT $V_{\mathrm{rf}}$ was increased at fixed $\Omega$ until the Mathieu stability parameter exceeded $q=0.908$, leading to the loss of ions from the LQT and the disappearance of the associated $m/Q$ peak from the ToF spectrum. This process was repeated for several values of $\Omega$ and for all $m/Q$ peaks in each ToF spectrum, leading to a list of $(\Omega,V_{\mathrm{rf}})$ pairs for each $m/Q$ peak. The $m/Q$ ratio for each peak was then found by fitting the $q=0.908$ stability parabola to the $(\Omega, V_{\mathrm{rf}})$ pairs with $m/Q$ as the only free parameter, as shown in the left panel of Fig. \ref{fig:calibration}.  The calculated $m/Q$ ratios for each peak are listed in Table \ref{tab:m}.   
\begin{table}[t]
    \caption{Comparison of actual and measured $m/Q$ ratios.}\label{tab:m}
   \centering
   \begin{tabular}{c|c|c|c}
   \hline
     Ion      & $m/Q_\mathrm{act}$ [amu/e]& $m/Q_\mathrm{meas}$ [amu/e]&Error\\
     \hline
      BaCl$^+$     & 172.78     & 174.43 & 1.0\%\\
      Ba$^+$       & 137.33  & 135.69 & 1.2\%\\
      SrCl$^+$    & 123.07  & 121.56 & 1.2\%\\
      Sr$^+$ & 87.62   &  84.95 & 2.7\%\\
      \hline
   \end{tabular}
\end{table}
The flight time of each peak was then mapped to its calculated $m/Q$ ratio which provides the flight time to $m/Q$ calibration curve shown in the right panel of Fig. \ref{fig:calibration}.  

To further characterize the device, a 3D Monte Carlo simulation of the LQT-ToF was performed using SIMION v8.1. In all simulations, we attempted to, at least partially, account for the Coulomb repulsion between the ions by simultaneously tracking groups of 1000 ions (a limit set by computational resources) and including their point-to-point and image-charge interactions. The results of these simulations are shown alongside data in Fig. \ref{fig:spectra} as dashed curves. Due to delays in the electronics and engineering imperfections, a $\sim1$~$\mu$s flight-time offset must be added to the simulations for maximum agreement. The simulated mass resolution, defined as $m/\Delta m=\frac{1}{2}t/ \Delta t$, where $t$ and $\Delta t$ are the mean and FWHM of the ions' flight-time distribution, was $\sim60$ whereas typical experimental mass resolution was found to be $\sim50$. This discrepancy might be explained by experimental imperfections such as misalignment and/or by the fact that the number of simulated ions may underestimate space-charge effects.

Finally, to determine the detection efficiency, the number of number of Ba$^+$ ions in the LQT prior to extraction was measured by observing laser-induced fluorescence. The Ba$^+$ laser system includes a $493$~nm cooling laser which drives the $^{1/2}$P$_{1/2}\leftarrow^{1/2}$S$_{1/2}$ transition ($\Gamma_{\mathrm{P}\leftarrow\mathrm{S}}=2\pi\times15$~MHz) and a $650$~nm repump laser which drives the $^{1/2}$P$_{1/2}\leftarrow^{1/2}$D$_{3/2}$ transition ($\Gamma_{\mathrm{P}\leftarrow\mathrm{D}}=2\pi\times5.6$~MHz).  These beams were both aligned along the axis of the LQT. Upon loading Ba$^+$ into the LQT, fluoresence from the $^{1/2}$P$_{1/2}\rightarrow^{1/2}$S$_{1/2}$ transition was measured by a photomultiplier tube while the cooling laser's frequency was scanned roughly $2\pi\times800$~MHz about the transition.   This signal was used to determine the number of trapped Ba$^+$ ions which was then compared to the number of Ba$^+$ ions detected by the ToF device.  The detection efficiency, which is sensitive to both ion-cloud and skimmer size, was found to be roughly $1\%$ which is in reasonable agreement with the Monte Carlo simulation prediction of $8\%$.   

\section{Proof-of-Principle Experiments}
To demonstrate the usefulness of this device for atomic, molecular, and chemical physics, we carried out two proof-of-principle experiments which are characteristic of atom-ion experiments performed with LQTs and hybrid atom-ion devices.  

\subsection{Photodissociation}

\begin{figure}[t]
\resizebox{\columnwidth}{!}{
    \includegraphics{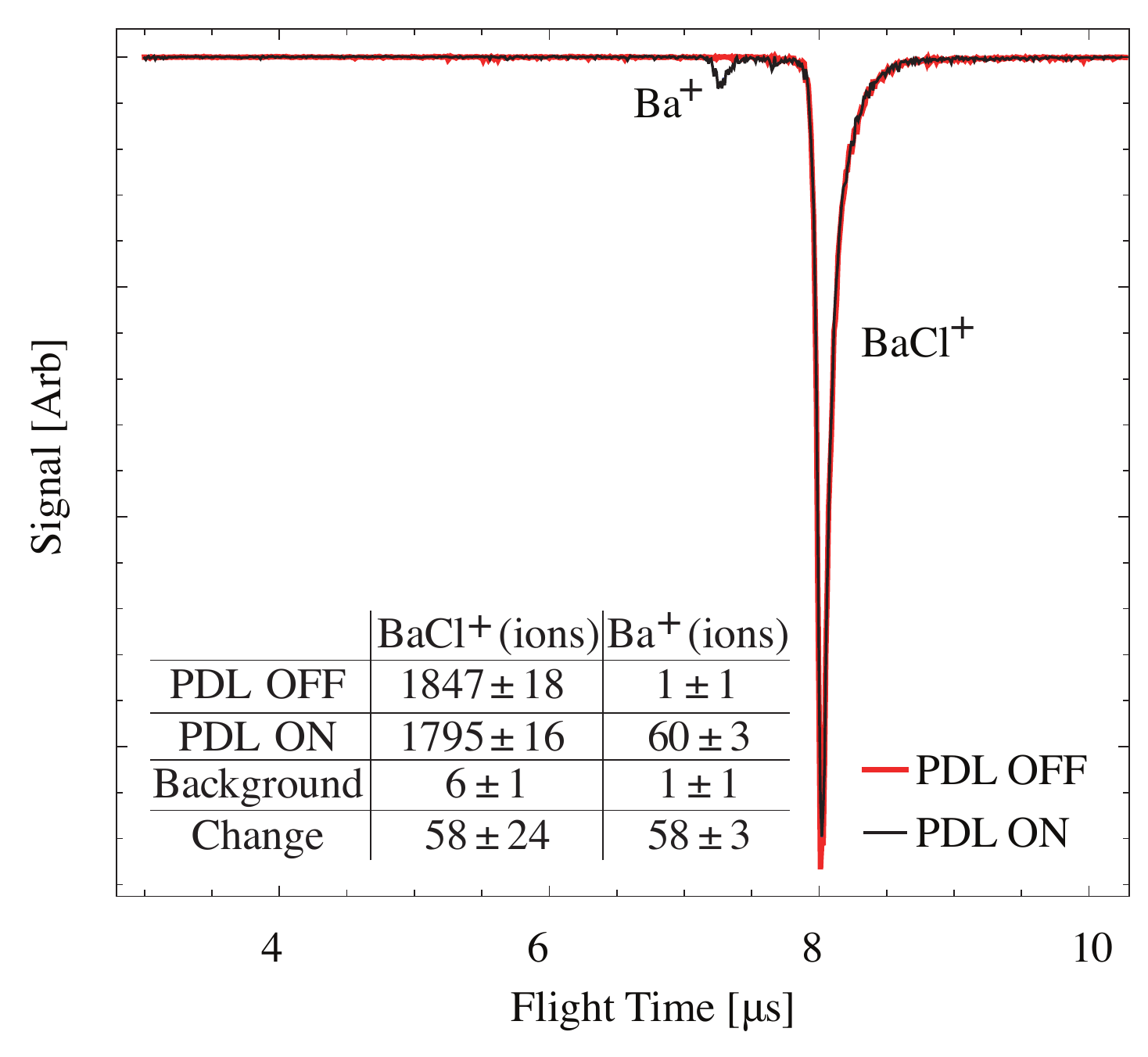}
}  \caption{(Color online) Experimental trace of the PDL's effect on a sample of trapped BaCl$^+$.  Photodissociation causes a loss in BaCl$^+$ which results in a slight increase in trapped Ba$^+$ in the ToF spectrum.    \label{fig:PD}}
\end{figure}

A major challenge to the present study of ultracold molecular ions is the current lack of spectroscopic information as discussed in Ref. \onlinecite{Saykally1981}. To advance this promising field, a new effort in molecular ion spectroscopy is needed. However, because fluorescence spectroscopy typically requires numbers of ions much larger than can be held in even a LQT, an alternate technique such as action spectroscopy, where spectroscopic photons are used to drive photofragmentation processes, must be employed. The LQT-ToF device could significantly aid this experimental effort as it affords high optical access and the ability to identify any action-spectroscopy ion products that are able to be co-trapped with the parent molecular ion. To prove the usefulness of this device for molecular-ion action spectroscopy, we performed a simple experiment to observe the photodissociation of BaCl$^+$, a candidate for ultracold molecular ion studies\cite{Hudson2009}.  This type of data is a necessary first step to understanding the structure for these relatively unstudied molecular ions. 

For this experiment, BaCl$^+$ and Ba$^+$ were first loaded into the LQT by ablating a BaCl$_2$ target.  The Ba$^+$ ions were removed from the trap using mass-filtering techniques\cite{Douglas2005}. Next, the BaCl$^+$ ions were exposed to laser pulses produced by a pulsed dye laser (PDL), which drive an electronic transition from the X$^1\Sigma^+$ state to the repulsive wall of the A$^1\Pi$ state resulting in the dissociation of BaCl$^+$ into Ba$^+$ (trapped) and Cl (untrapped) fragments\cite{Chen2011}.  To prove the sensitivity of the device, we used $10$ PDL pulses at $100$~$\mu$J/mm$^2$ per $10$~ns pulse and a photon energy of $\nu=41990$~cm$^{-1}$ which was red-detuned from the peak-resonance at $\nu_0\simeq46000$~cm$^{-1}$\cite{Chen2011}, and as a result, led to a small action-spectroscopy Ba$^+$ ion-product signal.  As seen in Fig. \ref{fig:PD}, the minimal discernible action-spectroscopy ion-product signal was measured to be approximately $1$~part~in~$1000$.  By detecting small increases in product ion from background levels, this method of fragment detection is much more sensitive than the previously-used method of trap depletion\cite{Chen2011}, in which small losses of parent ion are detected. This affords the ability to search for sensitive processes like predissociation, which are rovibronically state-selective and thus necessary for the ultracold-molecular-ion community, but typically have ion-product branching ratios of only one part in a few hundred.

\subsection{Chemical Reactions}
\begin{figure}[t]
\resizebox{\columnwidth}{!}{
    \includegraphics{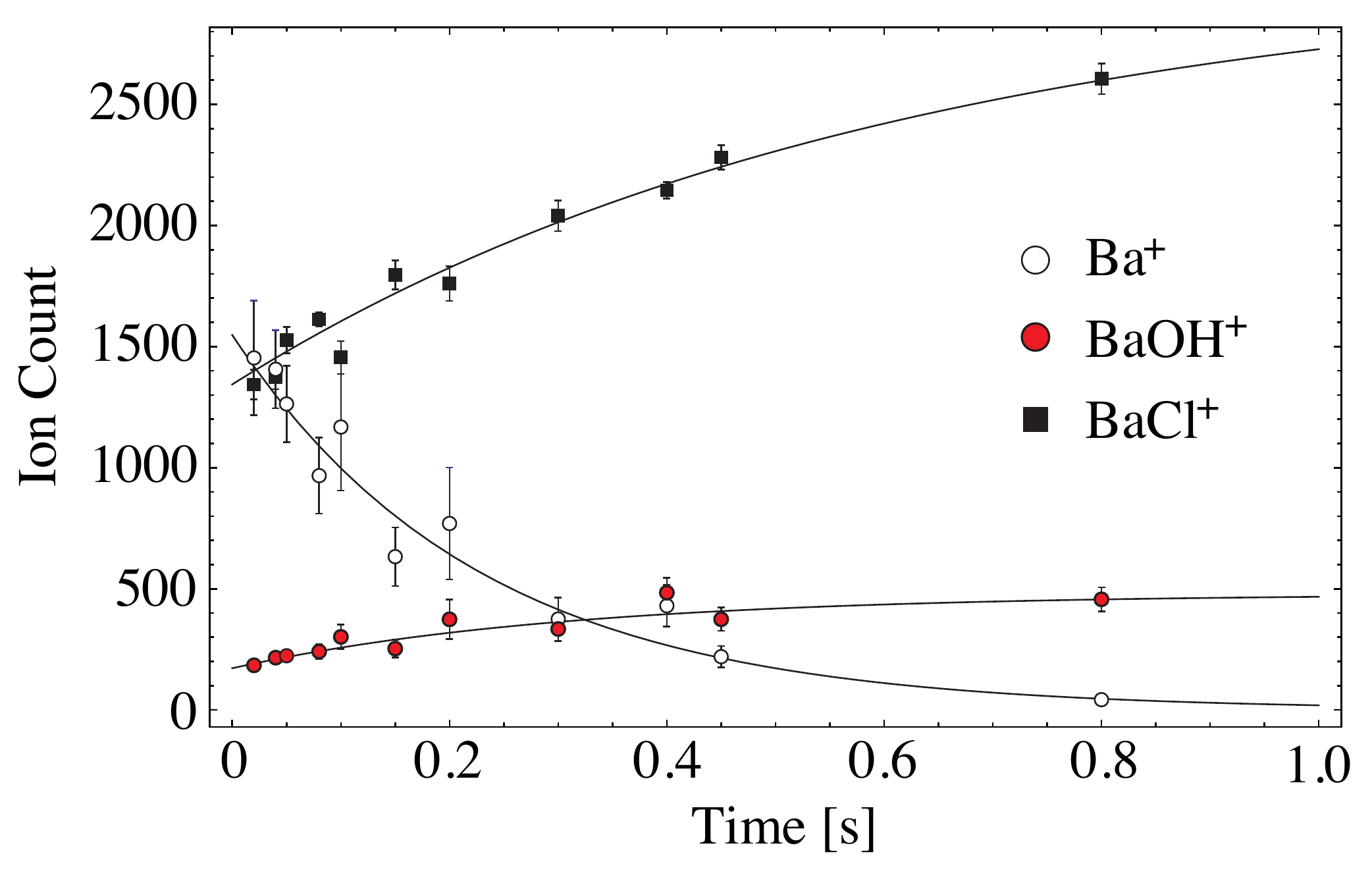}
}  \caption{(Color online) Background reactions between trapped Ba$^+$ and background gas resulting in the increase in trapped BaCl$^+$ and BaOH$^+$.  The reacting background gases are most likely H$_2$O ($2\times10^{-6}$~mbar) and HCl ($2\times10^{-7}$~mbar).  The total number of trapped ions remains $\sim3000$.\label{fig:reactions}}
\end{figure}

In the emerging field of hybrid atom-ion devices, quantum chemistry occurring between cold trapped ions and ultracold external neutral species is beginning to be explored\cite{Zipkes2010,Schmid2010,Rellergert2011,Hall2011,Grier2009}.  However, because these studies employ LQTs, it is often difficult to conclusively identify the products of these chemical reactions using current techniques, as aforementioned. These difficulties are especially poignant as there is some disagreement between theory and experiment in characterizing the branching ratio of radiative association to direct charge-transfer processes\cite{Rellergert2011,Zipkes2010a}.

To demonstrate the effectiveness of this device for these types of experiments, we performed a simple measurement to characterize the reaction rates between trapped Ba$^+$ and background gases which were present while the vacuum chamber was being pumped down to base pressure. In this experiment, Ba$^+$ and BaCl$^+$ were first loaded into the trap via ablation of the BaCl$_2$ target. These trapped ions were allowed to interact with an assortment of background gases for varying times.  The predominant gases were H$_2$O ($2\times10^{-6}$~mbar), HCl ($2\times10^{-7}$~mbar), and N$_2$ ($2\times10^{-7}$~mbar) as measured by a residual gas analyzer.  The presence of the HCl background gas can be explained by the presence of an AlCl$_3$ solid ablation target, which is known to react with H$_2$O to form HCl\cite{Elmboldt2012}. Following the interaction period between trapped ions and the background gas, the remaining ions were extracted from the trap and the ToF spectra recorded. Integration of the spectra for various trap times, $\tau<1$~s, led to the data shown in Fig. \ref{fig:reactions}. From this data it appears that trapped Ba$^+$ reacts with the background gas to create both BaCl$^+$ and BaOH$^+$\cite{Murad1982}.  The total ion count remains approximately $3000$ for all trap times, suggesting that this is a closed system and thus the most likely reactions are
\begin{align}
\mathrm{Ba}^{+}+\mathrm{HCl}&\rightarrow \mathrm{BaCl}^++\mathrm{H}\\
\mathrm{Ba}^{+}+\mathrm{H}_2\mathrm{O}&\rightarrow \mathrm{BaOH}^++\mathrm{H}.
\end{align}
By fitting exponentials to the data we determine the above reactions to have rates of $\Gamma_{\mathrm{(1)}}=1.7\pm0.1$~Hz and $\Gamma_{\mathrm{(2)}}=3.2\pm0.6$~Hz.  These rates are in good agreement with the $4.4\pm0.3$~Hz loss rate of Ba$^+$.  Estimates of the rate constants are $k_{\mathrm{(1)}}=3\times10^{-10}$~cm$^{3}$~s$^{-1}$ and $k_{\mathrm{(2)}}=6\times10^{-11}$~cm$^{3}$~s$^{-1}$, subject, of course, to the accuracy of the residual gas analyzer. From this data, it is clear that the LQT-ToF can be used to measure reaction rates to an accuracy of a few hundred mHz and, combined with the results from the previous section, can constrain the branching ratio of radiative association to direct charge-transfer processes to $1$~part~in~$1000$, representing an order of magnitude gain in sensitivity\cite{Rellergert2011}.

\section{Further Optimization}

The LQT-ToF device can be further optimized beyond its current implementation with the addition of several new features.  First, the current CEM can be substituted for a multichannel plate (MCP), which would increase the detection area and thus the detection efficiency of the device.  The operator control of the device would also be enhanced by the added ability to spatially resolve the ion beam incident upon the MCP.  Additionally, the use of a MCP would reduce the risk of signal saturation.  Second, cylindrical Einzel lenses can be used in place of the current spherical Einzel lenses to afford more flexibility and operator control in collimating the ion beam. Third, different skimmer geometries can be explored in order to deliver the optimal balance of detection efficiency and off-axis ion discrimination.  Rather than a skimmer with a circular aperture, as was used here, a skimmer with a narrow slit parallel to the LQT axis could be used to increase the number of ions which pass into the drift tube without sacrificing mass resolution.

\section{Conclusion}
We have demonstrated the implementation of a novel, simplified ToF mass spectrometer with medium mass resolution ($\sim50$), which is designed to couple directly to a LQT using radial extraction. To establish the usefulness of this device to atomic, molecular, and chemical physics we have performed two proof-of-principle experiments, which are germane to these fields of study.  First we used the LQT-ToF device to record an action spectroscopy signal with sufficient sensitivity to carry out \emph{e.g.} predissociation spectroscopy of molecular ions. In fact, we are currently using this device to search for a predicted predissociation channel in BaCl$^+$\cite{Chen2011}. Next, we demonstrated the ability of the device to characterize chemical reactions of trapped ionic species.  Here we showed significant improvements over current techniques in detection sensitivity, which should afford the ability to resolve discrepancies between experiment and theory regarding the branching ratio of radiative association to direct charge-transfer processes.  Finally, we have offered several simple improvements that could be implemented to improve the resolution and efficiency of the device.

\section{Acknowledgements}
The authors thank Paul Jordan from Jordan TOF Products, Inc. and Alexander Makarov from Thermo Fisher Scientific for their informative conversations on ToF mass spectrometry, Craig Taatjes for useful discussions involving ion-chemical reactions, and Shylo Stiteler for his help with the machining of the LQT-ToF device. This work was supported by NSF and ARO grants, PHY-0855683 and W911NF-10-1-0505.

\end{document}